\documentclass[review]{elsarticle}

\usepackage{lineno,hyperref}
\modulolinenumbers[5]

\usepackage[a4paper, body={16cm,22cm}]{geometry} 

\usepackage{amsmath,amssymb,amsfonts}
\usepackage{graphicx}
\usepackage{textcomp}
\usepackage{xcolor}
\def\BibTeX{{\rm B\kern-.05em{\sc i\kern-.025em b}\kern-.08em
		T\kern-.1667em\lower.7ex\hbox{E}\kern-.125emX}}

\usepackage{mathtools}  
\usepackage{amsmath}
\usepackage{amssymb}
\usepackage{tabulary}
\usepackage{booktabs}
\usepackage{amsmath}
\usepackage{algorithm}
\usepackage[noend]{algpseudocode}
\usepackage{balance}
\makeatletter
\def\BState{\State\hskip-\ALG@thistlm}

\makeatletter
\def\ps@pprintTitle{%
	\let\@oddhead\@empty
	\let\@evenhead\@empty
	\def\@oddfoot{\centerline{\thepage}}%
	\let\@evenfoot\@oddfoot}
\makeatother

\usepackage[T1]{fontenc}
\usepackage[latin9]{inputenc}
\usepackage{array}
\usepackage{float}
\usepackage{rotfloat}
\usepackage{booktabs}
\usepackage{calc}
\usepackage{amssymb}
\usepackage{graphicx}

\newcommand{\highlight}[1]{\textcolor{black}{#1}}

\usepackage[caption=false,font=footnotesize]{subfig}

\newcommand{\highlightnew}[1]{\textcolor{black}{#1}}

\begin{document}

\begin{frontmatter}
	

\title{Multiple Dynamic Pricing for Demand Response with Adaptive Clustering-based Customer Segmentation in Smart Grids}


\author[mymainaddress]{Fanlin~Meng\corref{mycorrespondingauthor}}
\cortext[mycorrespondingauthor]{Corresponding author}
\ead{fanlin.meng@essex.ac.uk}

\author[mysecondaryaddress]{Qian~Ma}
\ead{mqwilliam@hotmail.com}

\author[mysecondaryaddress]{Zixu~Liu}
\ead{zixu.liu@manchester.ac.uk} 

\author[mysecondaryaddress]{Xiao-Jun~Zeng}
\ead{x.zeng@manchester.ac.uk}

\address[mymainaddress]{Department of Mathematical Sciences, University of Essex, Colchester CO4 3SQ, UK}
\address[mysecondaryaddress]{Department of Computer Science, University of Manchester, Manchester M13 9PL, UK}

\date{}  

\begin{abstract}
	In this paper, we propose a realistic multiple dynamic pricing approach to demand response in the retail market. First, an adaptive clustering-based customer segmentation framework is proposed to categorize customers into different groups to enable the effective identification of usage patterns. Second, customized demand models with important market constraints which capture the price-demand relationship explicitly, are developed for each group of customers to improve the model accuracy and enable meaningful pricing. Third, the multiple pricing based demand response is formulated as a profit maximization problem subject to realistic market constraints. The overall aim of the proposed scalable and practical method aims to achieve `right' prices for `right' customers so as to benefit various stakeholders in the system such as grid operators, customers and retailers. The proposed multiple pricing framework is evaluated via simulations based on real-world datasets. 
\end{abstract}


\end{frontmatter}


\section{Introduction}
Dynamic pricing based demand response and demand side management \cite{batista2018demand, martinez2018optimizing} has attracted enormous attentions from both academia and industry in recent years. In particular, game-theoretic and machine learning based approaches (e.g., \cite{srinivasan2017game, zhang2017deep,bahrami2017online,lu2018dynamic, wang2020deep,
zhong2021deep}), which model the interactions between the energy retailer and customers, and study customers' energy consumption behaviours under dynamic pricing are emerging. However, most existing studies consider uniform dynamic pricing where customers are offered same prices.  Since customers have different consumption patterns and behaviours, uniform pricing might not be effective in capturing such differences among customers. For instance, a price which is a right incentive for price-sensitive customers might not be sensible for price-insensitive customers. 

An intuitive alternative is individual pricing where each customer receives a different price signal. For instance, \cite{hayes2017residential} adopted an individualized price policies in residential demand management where each customer receives a separate electricity pricing scheme in order to optimally manage flexible demands. However, individual pricing could result into a too complicated pricing optimization problem and suffer from scalability issues in practical applications. Moreover, it has to deal with individual highly random demand models, which is less meaningful for the customer relationship management. 

Multiple pricing, which relates to price discrimination and non-linear pricing \cite{wilson1993nonlinear}, is another pricing mechanism but so far has attracted only a very limited attentions in energy demand management \cite{simshauser2017price} \cite{tushar2017price}. However, multiple pricing is a fundamentally different and much more effective pricing mechanism  in comparison with the uniformed pricing. The reason is that a unformed price is, on the one hand, too high for some  customers whose willingness to pay  is lower than the price and leading to loss of this group of customers; on the other hand, too low for some other customers whose willingness to pay  is higher that the price and leading to loss of potential profits from this group of customers. The multiple pricing can successfully overcome such limitations, by offering the right prices to the right customers,  and enable retailers to reach all possible customers and realize all profit potential. Therefore introducing and investigating multiple pricing mechanism in energy demand  management is not just a simple extension from uniformed pricing to multiple pricing, but a significantly important research topic to enable energy retailers to better serve their customers and archive their full profit potential.  Along this direction, the existing few studies on multiple-pricing for demand response and smart grids can be grouped into three directions. 

For the first research direction, it presumes customer segments are given as a prior either represented by typical pre-known customer profiles or assuming customer segmentation has already been implemented beforehand. Therefore, the research mainly focuses on the multiple pricing design and optimization. For instance, \cite{yang2018framework} proposed a framework of customizing electricity retail prices based on pre-known customer segments where both the price values and structure were optimized. \cite{meng2017optimal} developed a bilevel optimization based multiple pricing approach to demand response where customer segmentation was assumed to have been done beforehand.

An important step to achieve efficient multiple pricing is customer segmentation \cite{flath2012cluster} where multiple or customized pricing can be designed to target different types of customers. Therefore, the second research direction attempts to identify customer segments directly from historical data based on clustering analysis, which can be used for multiple pricing design and optimization. For instance, \cite{haben2016analysis} proposed a finite mixture model based clustering approach to cluster residential customers based on smart meter data. To consider the price responsiveness of customers in the clustering process, \cite{asadinejad2016residential} proposed an approach to identify the price elasticity of each individual customer from historical consumption/ tariff and survey data. The identified price elasticities are adopted as attributes for the clustering-based customer segmentation to identify customer groups with distinct behaviours. 

The third research direction integrates the customer segmentation and multiple pricing within the same framework. For instance, \cite{kotouza2017segmentation} proposed a clustering-based approach to the customer segmentation for developing multiple pricing policy. \cite{chen2017Classification} proposed a electricity customer group classification approach based on K-means for customized price scheme design.  \cite{chrysopoulos2017customized} proposed a customized time-of-use pricing approach with Fuzzy C-Means for the customer segmentation where multi-objective particle swarm optimization was adopted for the pricing optimization. \cite{yang2018model} proposed a model of customizing electricity retail prices based on load profile clustering analysis where a density-based spatial clustering of applications with noise is first adopted to identify typical customer profiles directly from the customer consumption data, and then a customized ToU pricing optimization is implemented.

Although the above studies have provided valuable insights on how to design multiple pricing based on customer profiles or clustering analysis, there are still important research gaps to be filled and questions to be answered. Firstly, existing clustering-based customer segmentation approaches either directly work on individual price elasticities (computational expensive and difficult to derive and exhibit high randomness) or  electricity consumption data (fail to consider customers' price responsiveness thus less capable of getting desired customer segments). Instead, our proposed approach enables the practical and efficient use of energy consumption data and dynamic prices data simultaneously unlocked by the developed adaptive clustering-based customer segmentation. Secondly, different from existing studies which either assume pre-defined models for customers to reflect the price-demand relationship or model such a relationship implicitly using e.g. simulations, the proposed customized demand modelling approach models the price-demand relationship explicitly in a  data-driven manner, which provides fundamentals for the thereafter multiple pricing optimization.  In addition, key market constraints are included, which will ensure the demand model follows the market behaviour and enable meaningful and realistic multiple pricing decisions. Most importantly, the price-demand modelling is designed for each group of customers rather than for each individual customer, which carries two-fold benefits: 1) avoiding the high randomness in modelling each individual customer so as to improve the model accuracy; 2) ensuring the practicality of modelling in real-world application where there are usually hundreds of thousands customers being served by the same energy retailer. Thirdly, existing studies on multi-pricing optimization often assume pre-known mathematical models (e.g. formulated as optimization problems) for customer consumption behaviours whereas a systematic and integrated multiple pricing optimization framework connecting both the customer segmentation and the data-driven customized demand modelling is still missing. To this end,  in this paper we propose an integrated machine learning and optimization based multi-pricing framework with seamlessly connected key model components. The main contributions of the paper are summarized as follows. 

$\bullet$ Developed an integrated machine learning and optimization based multiple pricing framework, which connects adaptive clustering-based customer segmentation, customized demand modelling and multiple pricing optimization. It should be noted that the proposed data-driven customized demand modelling approach has high interpretability and can be directly embedded in the profit maximization problem of the retailer without affecting its mathematical characteristics as opposed to other grey/black box based models. The proposed framework also enables effective and scalable practical applications by transferring the pricing optimization problem of the retailer with up to multiple millions of customers to a problem with a few or dozen of customer segments. 

$\bullet$ Developed a fully data-driven and adaptive approach for clustering-based customer segmentation with customized demand modelling. The considered adaptive customer segmentation enables the efficient use of energy consumption data and dynamic prices data simultaneously to achieve desirable customer segments for the pricing purpose whilst the proposed customized demand modelling approach models the price-demand relationship explicitly in a data-driven manner with important and realistic market constraints included  to enable meaningful and effective pricing decisions. 

$\bullet$ The profit maximization problem is formulated as a multiple pricing optimization problem such that right prices are offered to right customers. The effectiveness of the proposed framework is further evaluated using real-world datasets. 

The remainder of this paper is organized as follows. The system framework is described in Section \ref{system framework}. The proposed data-driven and adaptive customer segmentation and customized demand modelling approaches are detailed in Section \ref{adpative-cluster}. The multiple pricing optimization model is given in Section \ref{multi-pricing}. Section \ref{simulations} presents the results and the paper is concluded in Section \ref{conclusion}.

\section{System Framework} \label{system framework}

\begin{figure}[!t]				
	\centering
	\includegraphics[width = 12cm]{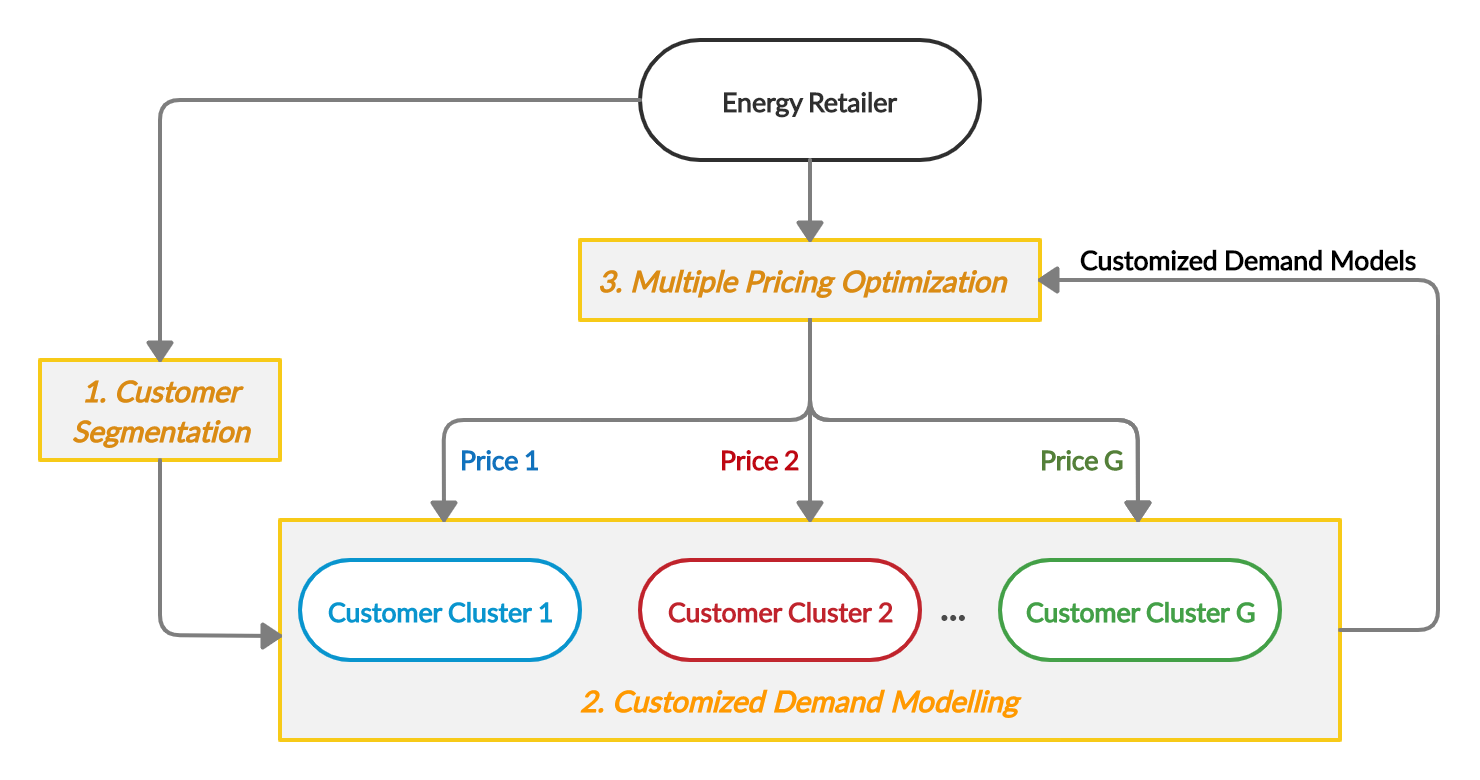}
	\caption{ The proposed multiple pricing framework.} 
	\label{figure:method_framework}
\end{figure}

As aforementioned, the proposed multiple pricing strategy can not only offer right prices to right customers but also has the great capability to help retailers realize the profit potential in the market due to its non-linear pricing characteristics. By leveraging the intrinsic incentives of retailers for innovative applications such as the proposed novel pricing strategy and demand response, it will in general benefit different stakeholders in the system such as customers, retailers and grid operators. Motivated by the above analysis and to further demonstrate the proposed pricing strategy, in this paper we propose a multiple pricing framework for the demand response management based on customer segmentation, which is illustrated in Figure \ref{figure:method_framework}. It consists of the following key components: 1) adaptive clustering-based customer segmentation; 2) customized price-demand modelling for each customer segment; and 3) the multiple pricing optimization. 

Suppose that the energy retailer purchases electricity from the wholesale market, reprices and sells it in the retail market to satisfy its business goals (e.g., short-term goals such as  profit maximization or long term goals such as market share)  under certain constraints/ factors (e.g. competitors' prices). To enable the multiple pricing framework in such a context, the energy retailer firstly implements customer segmentation based on historical electricity consumption and price data. Secondly, based on the clustering analysis, a customized data-driven price-demand modelling approach which captures the cross-relationship between demand and prices over different time periods is developed for each cluster of customers.  Thirdly, the customized price-demand models identified in the previous step are embedded into the multiple pricing optimization model to maximize the profit for the retailer subject to relevant market constraints. Finally, the optimal multiple prices are obtained to offer `right' prices to `right' customers.

\vspace{-8pt}

\section{ Adaptive Customer Segmentation with Customized Demand Modelling} \label{adpative-cluster}

In this section, we propose a dynamic and adaptive framework for customer segmentation with customized demand modelling. First, an adaptive clustering-based customer segmentation is developed to enable the efficient use of electricity consumption and dynamic price data. Due to behaviour changes of customers, the customer segmentation needs to be dynamically updated e.g.  monthly or seasonally. Following each customer segmentation, demand models for each customer group need more frequent updates (e.g. daily) with the new arrival of historical demand and price data. The updated demand models are then embedded into the pricing optimization to find optimal multiple prices (e.g. daily). In the following, the overall framework is firstly presented, followed by clustering details and the customized demand modelling. 

\subsection{Overall Framework} \label{subsec-overall-framework}

\begin{figure}[!t]				
	\centering
	\includegraphics[width=10cm]{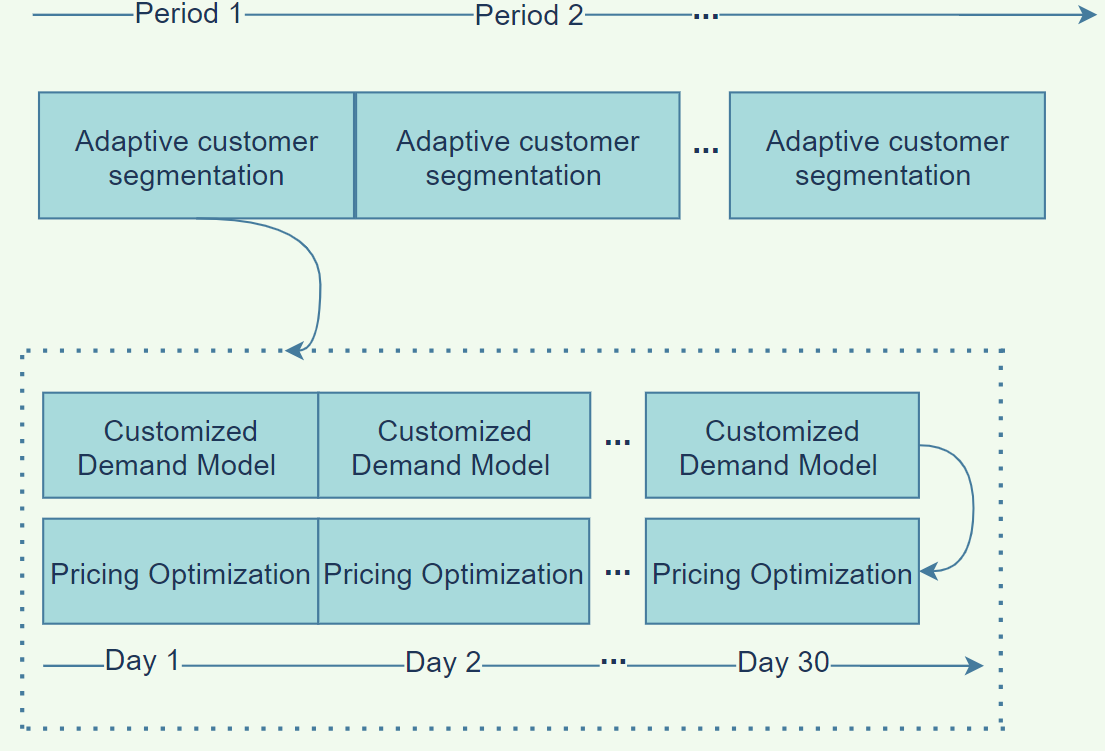}
	\caption{Dynamic and adaptive customer segmentation.} 
	\label{figure:apative_framework1}
\end{figure}

The overall framework and timescale of the proposed dynamic and adaptive customer segmentation with customized demand modelling is illustrated in Figure \ref{figure:apative_framework1} where the customer segmentation is updated monthly and the customized demand model and pricing optimization are updated daily.

\begin{figure}[!t]				
	\centering
	\includegraphics[width = 16cm]{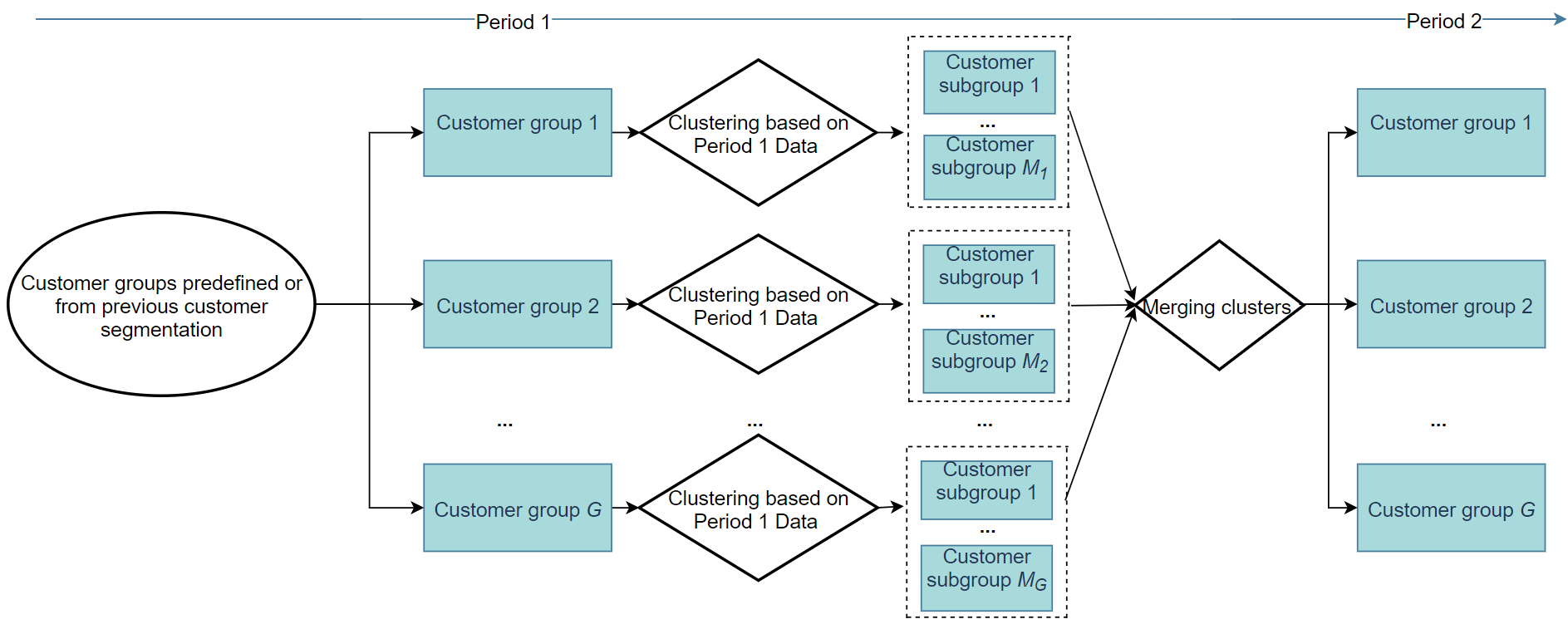}
	\caption{The proposed adaptive clustering-based customer segmentation.} 
	\label{figure:apative_framework2}
\end{figure}

The proposed adaptive clustering-based customer segmentation is illustrated in Figure \ref{figure:apative_framework2}, which consists of initial stage, sub-clustering stage and  cluster merging stage. As aforementioned, the proposed adaptive customer segmentation approach improves existing studies and enables the practical and efficient use of electricity consumption data and dynamic prices data simultaneously in clustering.

\subsubsection{Initial Stage} At the start of Period 1, based on the customer groups predefined or resulting from previous customer segmentation, different sets of tariffs are offered to different groups of customers during period 1 (e.g. one month). For each customer group, they share the same set of tariffs.

\subsubsection{Sub-clustering Stage}At the end of Period 1, based on newly obtained historical data (i.e. individual demand data under same set of dynamic prices) during Period 1, a sub-clustering is implemented for each group to get subgroups. 

\subsubsection{Cluster Merging Stage} Finally, a cluster merging stage is implemented to combine customer subgroups to the desired number of customer groups. This can be achieved by simply grouping together closest subgroup/ sub-cluster centroids. Alternatively, given that sub-clusters are aggregated electricity consumption data and the number of sub-clusters are usually small, customized demand models can be built for each sub-cluster based on the method proposed in subsection \ref{group demand modelling} to better capture the price-demand relationship for each subgroup. Then demand models with similar behaviours can be grouped together (e.g. through clustering coefficients of demand models ) to achieve the cluster merging. The newly updated customer groups are then used in Period 2 for rebuilding demand models and pricing optimization.  The same customer segmentation process repeats at the end of Period 2 when the new  historical data during Period 2 become available. 

The benefits of such adaptive clustering based customer segmentation are multifold: 1) compared with clustering all customers directly, it can overcome the difficulty of using demand data under different sets of dynamic prices (for different customer groups) in clustering and also avoid the random, error-prone and computational extensive process of obtaining individual-level/ household-level price elasticities or demand models; 2) the number of clusters in the sub-clustering stage and cluster merging stage can be easily adjusted based on practical application requirements to achieve a good balance of computational efficiency and accuracy. 

\subsection{Clustering Details}

For each time slot $h \in \mathcal{H} =\{1,2,...H\}$ on  past day $d \in \mathcal{D} =\{1,2,...D\}$, the historical price and consumption data of each customer $n \in \mathcal{N} =\{1,2,...N\}$ are assumed to be available and denoted as $p^h (d)$ and $y_n^h (d)$ respectively where $H$ represents the pricing optimization horizon (e.g, $H$ =24 indicates the pricing for each hour of the day), $D$ represents the total number of days in the historic data, and $N$ represents the total number of considered customers. Although the clustering details described in this subsection correspond directly to the sub-clustering stage in subsection \ref{subsec-overall-framework}, they are also readily available for the cluster merging stage subject to minimum adaptations (e.g. using subgroup demand model coefficients as clustering attributes). 

\subsubsection{Attributes selection}The clustering-based customer segmentation utilizes the energy consumption data of individual customer as input attributes of clustering algorithms. As aforementioned, some existing studies utilize individual customers' price elasticities as input attributes. However, such information is either very difficult to obtain or possess high randomness in the process of obtaining caused by high variations of usage patterns of each individual customer due to different life styles and behaviour uncertainties. In additional, the process is computationally extensive considering the number of individual customers in practice.  As a result, in this paper we utilize the electricity consumption data under dynamic pricing directly in the sub-clustering stage of the proposed adaptive clustering-based customer segmentation framework, which will avoid the randomness and additional uncertainty/ inaccuracy in the process of obtaining such price elasticity information of individual customers.

\subsubsection{Normalization}Due to that the segmentation aims to provide decision support for the thereafter customized demand modelling and multiple pricing optimization, our clustering aims to group together customers that respond similarly to the price signals. However, as different customers have different consumption levels, the absolute change of demand in response to the change of price will fluctuate very differently. For instance, for customers with relatively high energy consumptions, the absolute change in demand responding to price signals might look high, however, the relevant change (e.g. against his average consumption) might be low. Instead, a customer with relatively low energy demand level might have already changed his behaviour significantly in response to price signals even though the absolute change in demand is lower than customers with high consumption level. As in the demand modelling and pricing optimization, we are more interested in the relative responsiveness of demand to prices so as to design customized pricing strategy for each customer group, we need to normalize the consumption profiles of customers before adopting them as clustering attributes. In addition, we are interested in the demand changes in some particular time periods such as the peak demand time (early evenings) and off-peak demand time (night) to better reflect the demand shifting and demand reduction behaviours. Based on the above considerations, we normalize customers' consumption data against their daily average consumptions.

\subsubsection{Clustering algorithms}Assume we have $N$ observations (number of customers) denoted as $ \mathrm{\textbf{atr}_{1}}, \mathrm{\textbf{atr}_{2}}, ... \mathrm{\textbf{atr}_{N}}$ for the clustering algorithm. Each observation is a $r$-dimensional real vector, i.e. $\mathrm{\textbf{atr}_{n}} \in \mathbb{R}^r$ where $r$ represents the length of normalized electricity consumption data. \highlight{More specifically, for hourly dynamic pricing (i.e. $H=24$), the clustering attributes could be selected to be  historical hourly consumption data under dynamic pricing  for past $D$ days, i.e.  $\mathrm{\textbf{atr}_{n}} = \{y_n^1 (1),y_n^2 (1), ...y_n^{24} (1),...y_n^h (d), ...y_n^{24} (D)\}$. That is,  there are in total $24 \times D$ clustering attributes. For time-of-use pricing with three price segments (i.e.  $H=3$), the clustering attributes are historical average hourly consumption in each price segment for past $D$ days with a total of $3\times D$ attributes.}

\highlightnew{We demonstrate below the clustering procedure using the popular $K$-means, which can be easily adapted and applied to other clustering algorithms, e.g. Hierarchical Clustering \cite{ward1963hierarchical}. }

The optimal number of customer clusters is denoted as $ K$. Therefore, the $K$-means clustering aims to optimally assign each customer to a cluster to minimize the within-cluster variance, i.e. find the optimal partition sets for those $N$ observations $\mathrm{\textbf{C}} = \{C_1, C_2, ..., C_K\}$. Mathematically, the $K$-means is formulated as the following optimization problem: \begin{equation}
	\min_{\mathrm{\textbf{C}}} ~ \sum_{j=1}^K \sum_{\overline{\mathrm{\textbf{atr}}} \in C_j} d( \overline{\mathrm{\textbf{atr}}} , \mu_j) 
\end{equation}
where $ C_{j} $ is the set of observations  that correspond to cluster $ j $,  $\mu_j$ is the mean of observations in set $C_j$, which is also a $r$-dimensional real vector. Further, $\overline{\mathrm{\textbf{atr}}}$ represents each possible observation in set $C_j$.

The distance function $ d( \overline{\mathrm{\textbf{atr}}} , \mu_j) $	in our considered $K$-means is based on the Euclidean distance and is defined as follows: 
\begin{equation}
	\label{k-means}
	d( \overline{\mathrm{\textbf{atr}}} , \mu_j) =\sqrt{ \sum _{i=1}^{r} \left( \overline{atr^i}- {\mu}_j^i\right) ^{2}} 
\end{equation}
where $\overline{atr^i}$ is the $i$-th element in the vector $\overline{\mathrm{\textbf{atr}}}$ and ${\mu}_j^i$ is the $i$-th element in the vector  $\boldmath{\mu}_j$.

\subsubsection{Clustering Evaluation} 

Due to the lack of ground truth information, we consider internal clustering validation.  In this paper, we choose two internal evaluation metrics: the Silhouette Coefficient (SC) and the Davies-Bouldin index (DBI) \cite{de97dynamic}. SC is obtained by contrasting the average distance to objects in other clusters with the average distance of objects within the same cluster. That is, clusters with low similarity between clusters and high similarity within a cluster will have higher SC values. DBI is related to the ratio of the intra-cluster distances to the inter-cluster distances. Therefore, a lower DBI value indicates a better clustering configuration.  As can be seen from the above, these two metrics simultaneously consider not only within cluster distance but also the distance between clusters. That is, they measure both the intra-cluster and inter-cluster similarities. The optimal clustering algorithm and the number of clusters are chosen such that the above evaluation metrics are optimized.

\subsection{Customized Demand Modelling} \label{group demand modelling}

We denote the set of customer groups as $\mathcal{G} = \{1,2, ...G\}$.  To offer different pricing strategies to different customer groups, we need to build a customized demand modelling for each group $g\in \mathcal{G}$. Note that the proposed approach is fundamentally different from the individual-level approach which is computationally expensive and usually has a poor modelling accuracy due to the loss of mutual compensation of usage variations between customers.  Instead, our proposed approach reduces the uncertainty and variation of individual usage behaviour via grouping customers with similar behaviour, which not only improves the modelling accuracy but also achieves great simplicity in computing. In addition, the proposed approach is able to distinguish amongst different usage behaviours of customer groups and enable the demand model that is appropriate to a given customer segment.

As mentioned in subsection \ref{subsec-overall-framework},  suppose we have customized demand models for each sub-cluster in the cluster merging stage already identified based on the below method, the customized demand modelling for each customer group after cluster merging can be achieved by aggregating demand models of relevant sub-clusters.

\subsubsection{Price-demand modelling} we adopt a linear price-demand modelling approach in this study with important market constraints to ensure the resulting demand model exhibits reasonable and rational market behaviours, which are important for the multiple pricing optimization.

The estimated demand function  for each group $ g \in \mathcal{G}$ during each hour $ h \in \mathcal{H} \triangleq  \{1,...,H \}$ can be represented as follows: \begin{equation}
	R_{h}^{g} \left(p^{1}, p^{2},...p^{H} \right) = \alpha _{h}^{g} + \beta _{h,1}^{g} p^{1}+ ...+ \beta _{h,H}^{g}p^{H}
\end{equation} where parameter $ \beta _{h,h}^{g} $ can be used to account for the self-elasticity or direct elasticity of customer group $g$ that measures the responsiveness of electricity demand at hour $h$ to changes in the electricity price at $h$. $ \beta _{h,l}^{g} $ are used to represent the cross-price elasticity of customer group $g$, which measures the responsiveness of the demand for the electricity at hour $h$ to changes in price at some other hour $l \neq h$. 

\subsubsection{Realistic market constraints} 

When the price at hour $h$ increases but prices at other times remain unchanged, the demand at hour $h$ usually decreases, and thus $ \beta _{h,h}^{g} $ is non-positive (see \eqref{elasity1}).  When the price of electricity at hour  $l$ increases but prices at other times remain unchanged, some demand at hour $l$ can be shifted to hour $h$, and therefore the demand at hour $h$ usually increases. Thus, the cross elasticity is non-negative (see \eqref{elasity2}).
\begin{align}
	&\beta_{h,h}^g \leq 0.  \label{elasity1} \\
	&\beta_{h,l}^g \geq 0 \mbox{ if }  h \neq l.\label{elasity2} 
\end{align}	

In addition to constraints Eqs. \eqref{elasity1}\eqref{elasity2}, we develop an important necessary and sufficient condition (see \eqref{elasity4}) for the electricity to be a demand consistent retail product to ensure the demand model follows a normal market behavior. 
\begin{align}
	&\beta_{h,h}^g +  \sum_{l \in \mathcal{H}, l \neq h} \beta_{l,h}^g \leq 0. \label{elasity4}
\end{align}

\highlight{To be more specific, for a retail product such as electricity, suppose there are several competitive prices  $P= \{p^{1}, p^{2},...p^{H}\}$ for this product and their corresponding demands are $Y=\{y^1, y^2,...y^H\}$. It is said that the given product in the market is demand consistent if for any two sets of prices $P_1= \{p^{1}_1, p^{2}_1,...p^{H}_1\}$ and	$P_2= \{p^{1}_2, p^{2}_2,...p^{H}_2\}$ satisfying $P_1 \leq P_2$ (defined as $p_1^{h} \leq p_2^{h} \ \forall h \in \{1,2...H\}$) the corresponding demands $\{y^1_1, y^2_1,...y^H_1\}$ and $\{y^1_2, y^2_2,...y^H_2\}$ satisfy $Y_1 = y^1_1 + y^2_1+...+y^H_1 \geq  Y_2 = y^1_2 + y^2_2+...+y^H_2$. In other words, when some prices are increased whereas other prices remains unchanged (that is, the overall market price is increased), the total market demand $Y= y_1+ y_2+ ... +y_H$ will decrease or remain unchanged. This market consistency condition is particularly important for the pricing purpose to  ensure the result of the pricing optimization being realistic. If such a condition is not satisfied, it may end up with an unrealistic optimal price decision where the overall market price increases, and the demand increases simultaneously. It would create unrealistic or non-existing profit as a result of the inaccurate demand model, which might eventually lead to wrong pricing/ market decisions by energy retailers. }

\subsubsection{Model estimation}Given the historical hourly demand and price  data for each customer group over the past $D$ days, the unknown coefficients in the above demand models can be identified by solving the following weighted least square optimization problem with constraints for each customer group:
\begin{equation} \label{qp_whole_demand}
	\begin{array}{lll}
		\min \sum_{d \in \mathcal{D}}\sum_{h \in \mathcal{H}} \lambda_g ^{D-d} (  \alpha _{h}^{g}+ \beta _{h,1}^{g}p^{1}(d)+...+    \beta _{h,H}^{g}p^{H}(d) - y_g^{h}(d) ) ^{2} \\
		\mbox{ subject to \eqref{elasity1}, \eqref{elasity2} and \eqref{elasity4}.}
	\end{array}
\end{equation}  
where $ 0< \lambda_g  \leq 1 $ is the forgetting factor that exponentially reduces the influence of old data. In the current form of the problem, it is difficult to solve using traditional mathematical methods such as least square (difficult in handling parameter $\lambda$ ) or recursive least square (difficulty in handling constraint \eqref{elasity4}). Instead, we reformulate the problem into a typical quadratic programming problem, which can be solved using off-the-shelf mathematical optimization packages.

\textbf{Remark 1.}
This customized demand modelling approach has desirable capabilities and characteristics such as higher demand modelling accuracy compared with the aggregated or individual demand modelling approaches. In addition, the model has high interpretability and can be directly embedded in the profit maximization problem of the retailer without affecting its mathematical characteristics (i.e. the resulting profit maximization problem remains to be well-defined and could be solved by conventional mathematical programming methods) as opposed to other grey/black box based demand models (e.g. neural networks) where the resulting profit maximization problem is likely to be ill-defined and usually needs to employ meta-heuristic methods. 

\section{Multiple pricing optimization} \label{multi-pricing}

As aforementioned, by offering `right' prices to `right' customers, it can not only improve the customer satisfaction (which is useful for retailers to keep or improve their market share) but also increase retailer's profit (which can be seen as an innovation service to reduce the cost). Specifically, the multiple pricing offers customized prices to satisfy needs of different customers. By offering `right' prices, it will also greatly incentivize customers to shave their peak demand, which in turn reduces the power generation cost (and therefore retailer's purchase cost).

With the customized demand modelling in place, we formulate the multiple pricing problem to maximize the profit for the retailer. By defining the minimum and maximum prices that the retailer can offer to each customer group, we have
\begin{equation} \label{price_constraints}
	\begin{array}{lll}
		p_{g,h}^{min} \leq p_{g}^{h} \leq p_{g,h}^{max}, \forall h \in \mathcal{H},~ \forall g \in  \mathcal{G}
	\end{array}
\end{equation}
\highlight{where $p_{g,h}^{min}$ and $p_{g,h}^{max}$ are usually set based on considerations such as the cost of wholesale electricity, customers' willingness and affordability to pay, and political pressure. On the one hand, to avoid a loss, the retail price is often set higher than the wholesale price. On the other hand, an upper bound of the retail price is usually imposed to create a reasonable long-term price image to withhold market share from competitors. In addition, government policy and customers' acceptability also force prices to be upper bounded.}

Due to the retail market regulation, a revenue cap $RE^{max}$ is usually in place for the energy retailer to ensure a sufficient number of low price periods and to improve customers' acceptability of pricing strategies. Therefore, we consider:
\begin{equation} \label{revenue_constraint}
	\begin{array}{lll}
		\sum _{g \in  \mathcal{G}} \sum _{h \in  \mathcal{H}} R_{h}^{g} p_{g}^{h}  \leq RE^{max}
	\end{array}
\end{equation} where $R_{h}^{g} =  \alpha_{h}^{g}+ \beta _{h,1}^{g}p_{g}^{1}+ ... + \beta _{h,H}^{g}p_{g}^{H}$.

\highlight{It should be noted that the above revenue constraint is important in practice from the pricing optimization perspective to ensure a sensible pricing strategy for the retailer. Since the electricity is a basic necessity and therefore less elastic, without such a constraint, a retailer could lift its profit greatly by increasing its prices aggressively.}

To have a good price image among customers (which is important for the energy retailer's market share in the long-term), we include a constraint such that the average prices offered by the retailer to each customer group are the same. This can be done by setting the average price offered to each customer group to be some predefined value denoted as $FP$ (e.g., the equivalent flat price). For instance, \cite{zugno2013bilevel} has included such kind of constraint in their uniform pricing model. Note that constraint \eqref{average price} can have similar effects as constraint  \eqref{revenue_constraint} in ensuring sufficient low price periods for customers. 
\begin{equation} \label{average price}
	\begin{array}{lll}
		\frac{\sum_{h\in \mathcal{H}} p_{g}^{h}}{H} = FP, ~ \forall g \in  \{1, 2,..., G\}
	\end{array}
\end{equation}
Finally, the pricing optimization problem for the retailer can be represented as the following profit maximization problem: 
\begin{equation} \label{profit maximization1}
	\begin{array}{ccc}
		\max \sum _{g \in G}^{} \sum _{h \in \mathcal{H}}   \left(p_{g}^{h}-c^{h} \right)  \times  R_{h}^{g} ( p_{g}^{1},  p_{g}^{2},... p_{g}^{H})  \\
		\mbox{subject to constraints (\ref{price_constraints}),  (\ref{revenue_constraint}) or/ and \eqref{average price}}
	\end{array}
\end{equation} where $c^h$ is the unit supply cost of the retailer. 

The above problem is a quadratic constrained quadratic programming (QCQP) problem when constraint \eqref{revenue_constraint} is included. Otherwise, it is a  standard quadratic programming (QP) problem. Both can be solved using the off-the-shelf mathematical optimization packages.

\noindent\textbf{Remark 2.}	
The considered multiple pricing framework with customer segmentation and customized demand modelling has the capability of customizing different pricing strategies for different group of customers and  offers the following benefits: 1) it ensures the practicability and tractability of the multiple pricing model in real world applications where there could be millions of customers served by the same energy retailer;  2) it could enhance
the modelling accuracy by customized price-demand modelling where the high randomness of individual customers could cancel out with each other within the group compared with the individual demand modelling approach required for individual/ personalized pricing schemes. 

\section{Results}\label{simulations}

In this section, we first use a real-world smart meter dataset to demonstrate the effectiveness of the proposed adaptive customer segmentation framework. Then, results for the multiple pricing optimization are presented. 

\subsection{Residential Customers Clustering Analysis}

\subsubsection{Dataset}

\highlightnew{The smart meter dataset used in this paper is from the Irish smart meter trial, which is available at \cite{cer_smart_meter}. The dataset include half-hourly energy consumption from residential smart meters from July 2009 to the end of 2010. All customers were placed under flat prices from July 2009 till the end of 2009. They were divided into different groups and offered different time-of-use (TOU) tariffs with different levels of peak, day and off-peak prices from January 2010. }

\subsubsection{Clustering Results}

To demonstrate the adaptive clustering based customer segmentation framework and capture price changes between flat prices and TOU, we consider smart meter data of customers under four residential TOU tariffs (Tariffs A, B, C, D- termed as TA, TB, TC, TD) \cite{cer_smart_meter} during the time period between December 01, 2009 and January 31, 2010. After the data preprocessing, we have obtained a smart meter dataset of 3208 customers.  

\begin{figure}[!t]
	\centering{}
	\subfloat[Cluster 1] {\centering{}\includegraphics[width=8cm]{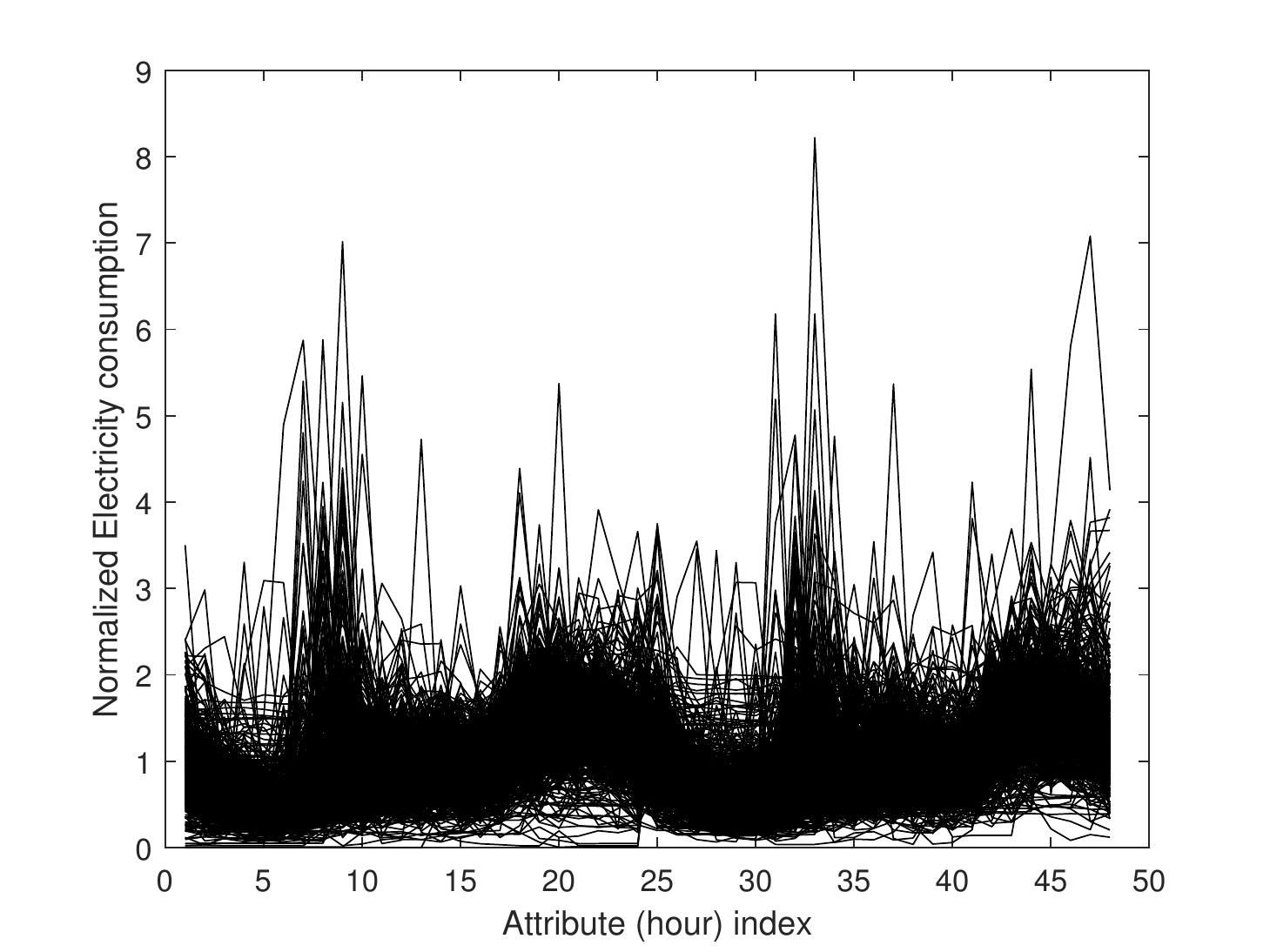}}
	\subfloat[Cluster 2]{\centering{}\includegraphics[width=8cm]{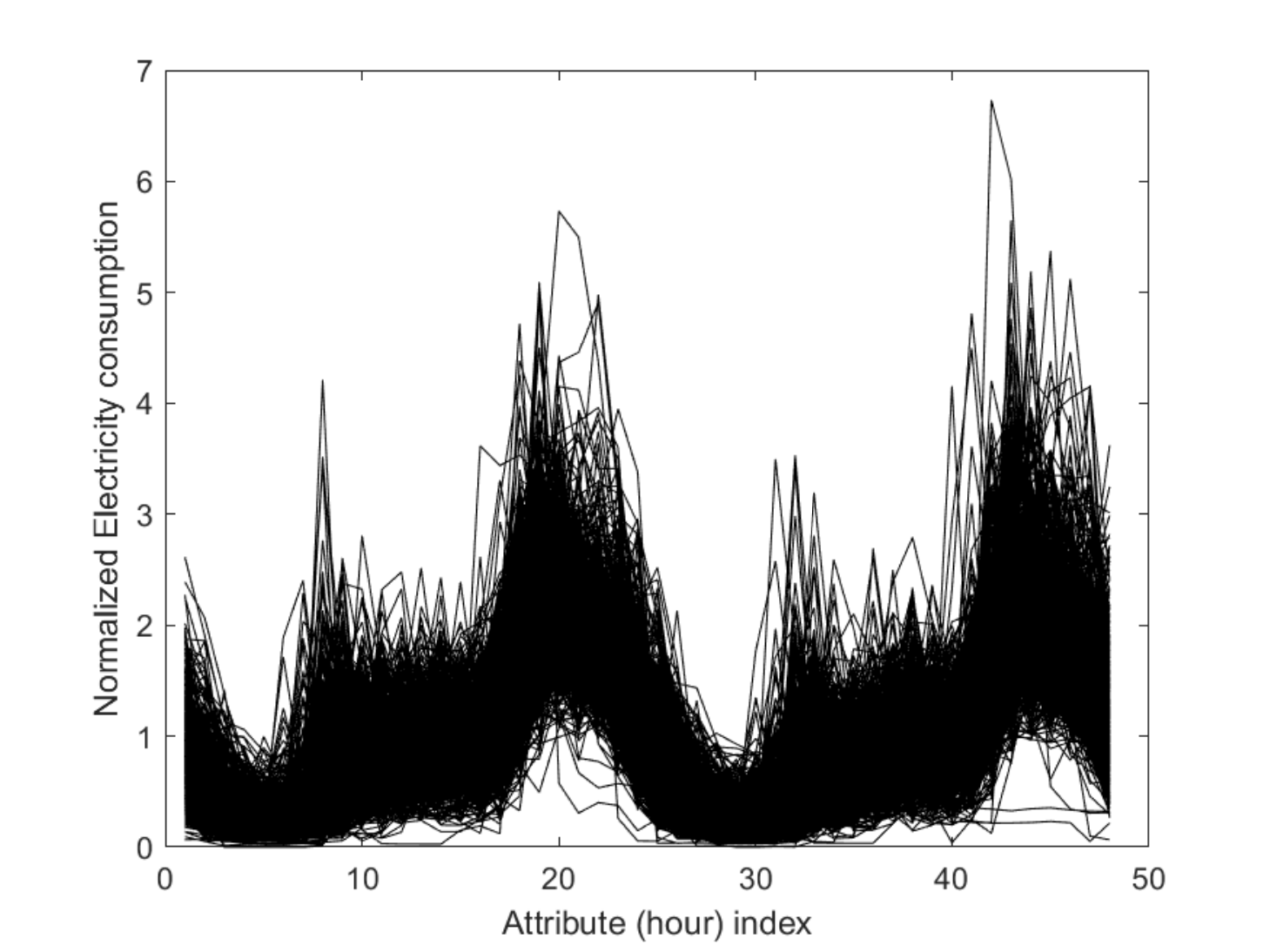}}\vfill{} 
	\subfloat[Cluster 3]{\centering{}\includegraphics[width=8cm]{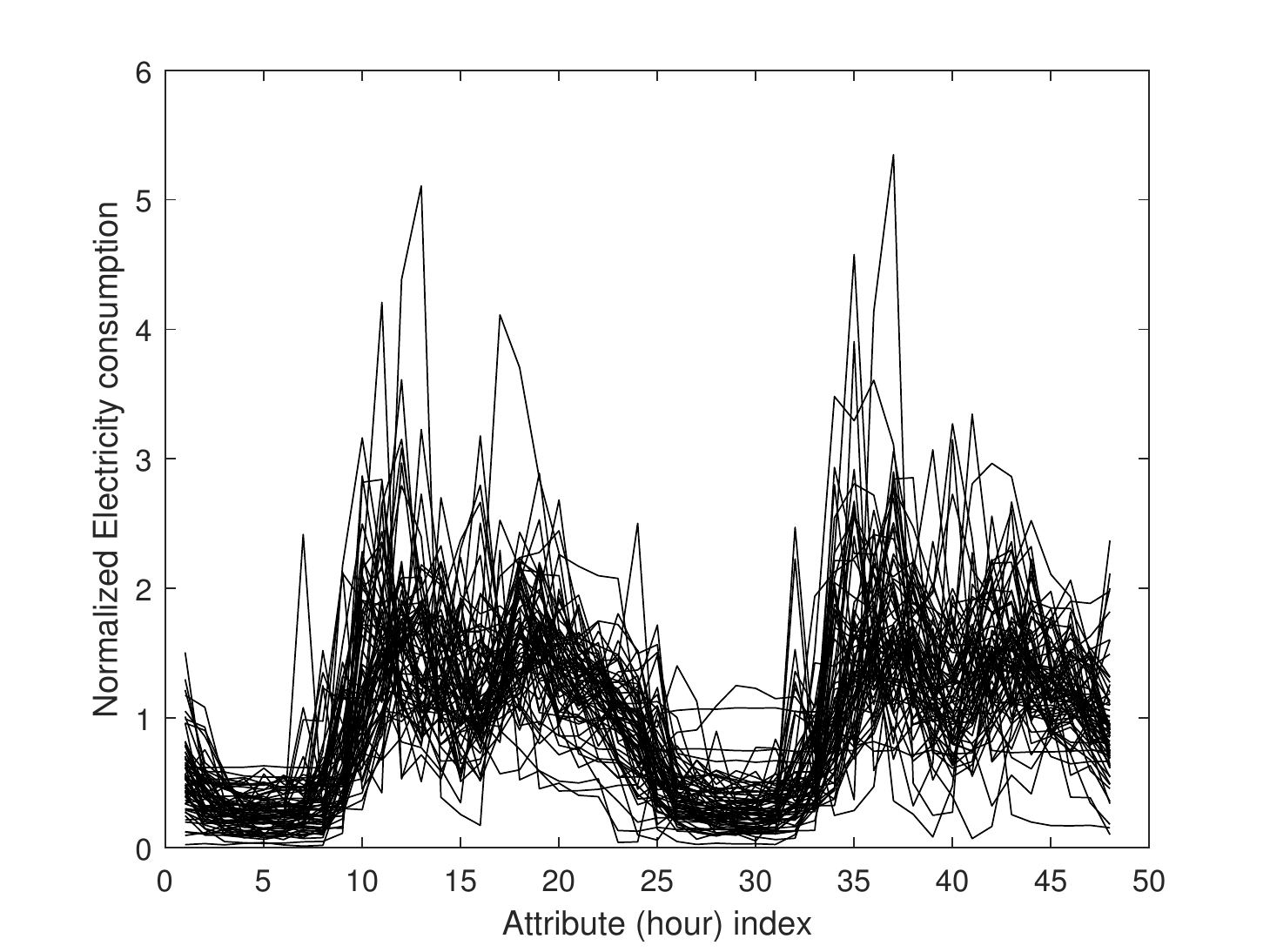}} \subfloat[Cluster 4]{\centering{}\includegraphics[width=8cm]{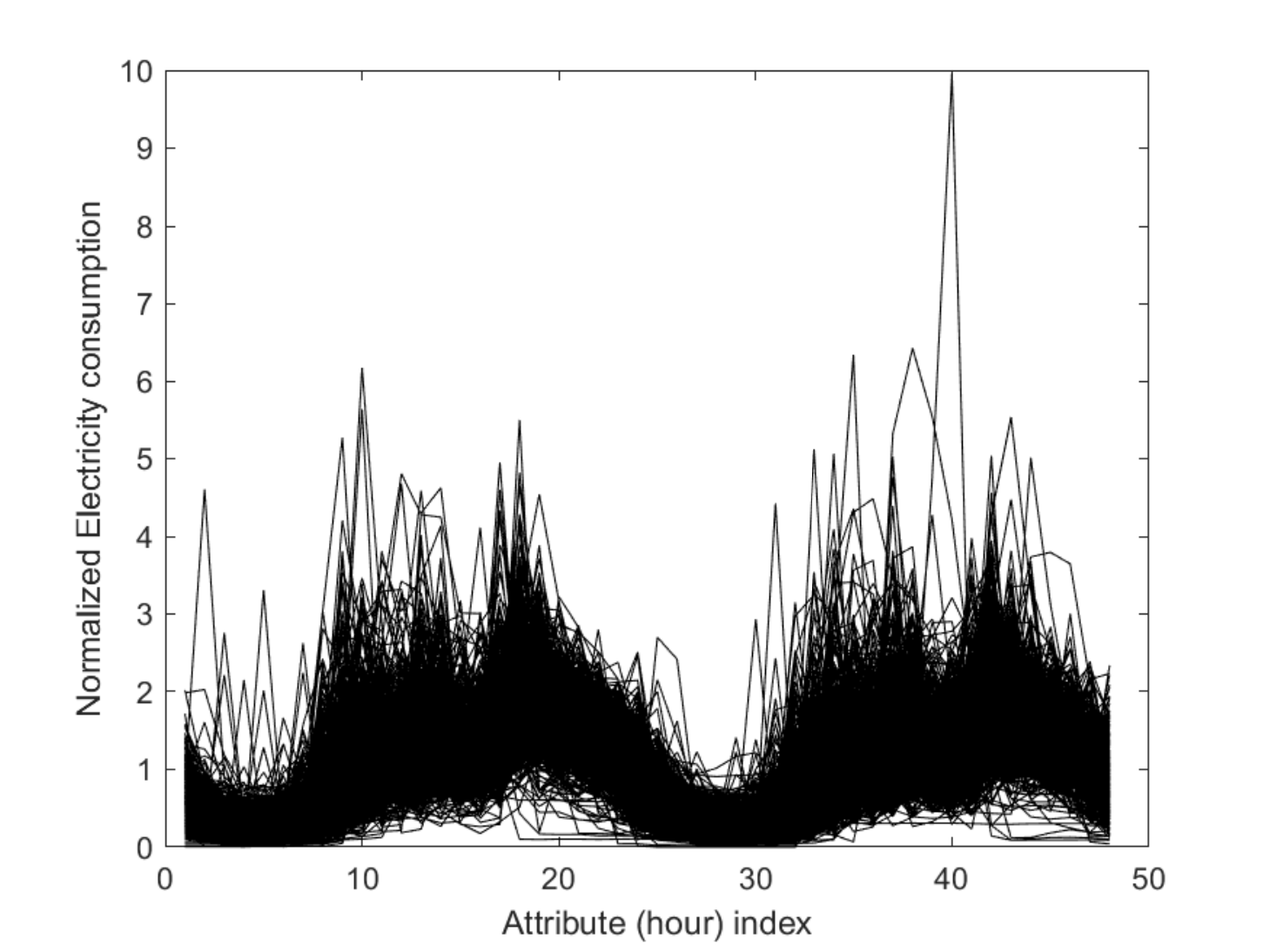}}
	\caption{\label{fig:finalclusterconsumption}The normalized consumption profiles of each cluster.}
\end{figure}

\begin{figure}[!t]
	\centering{}
	\includegraphics[width=12cm]{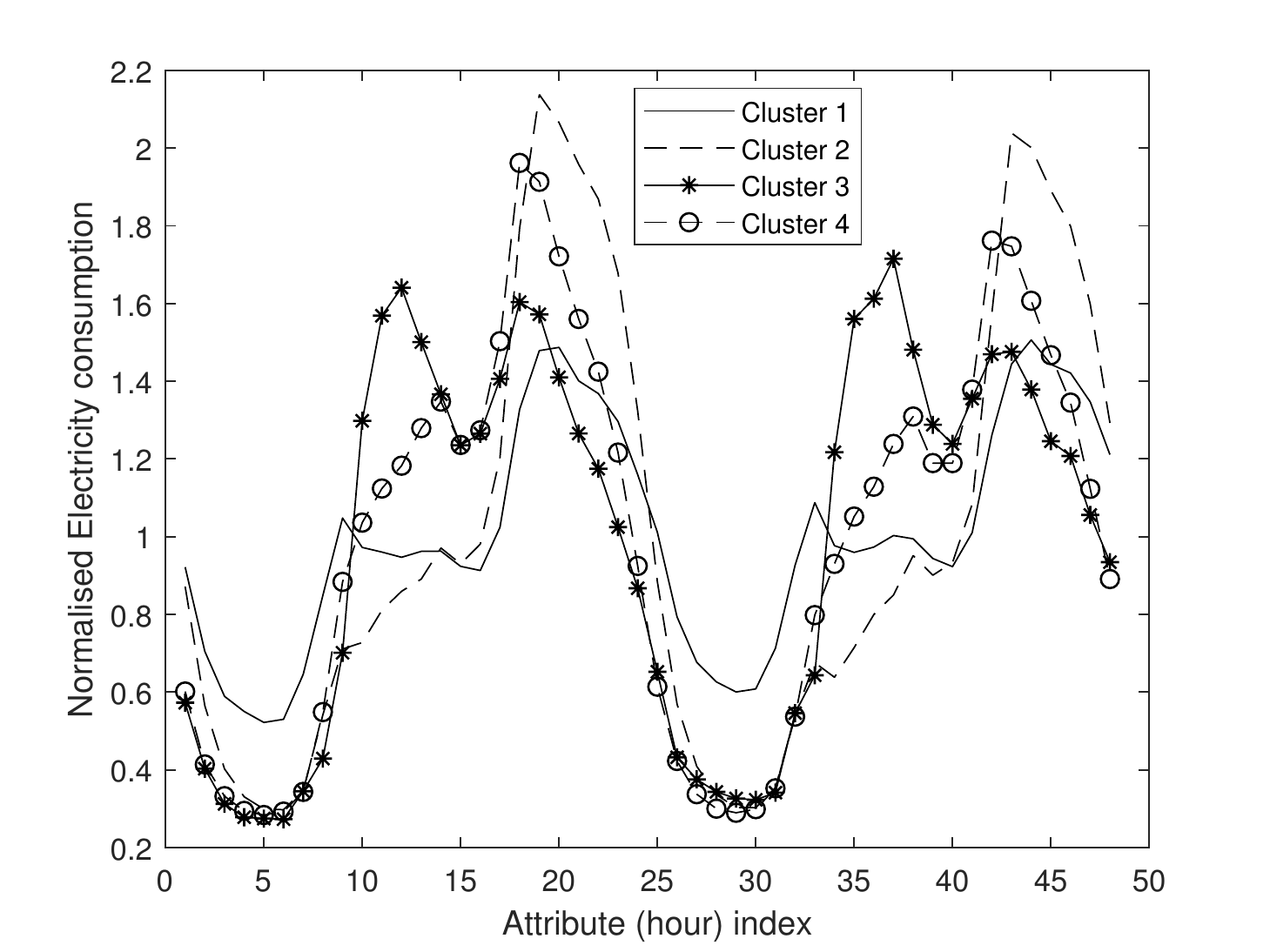}
	\caption{\label{fig:clustercentroids} Electricity consumption of cluster centroids.}
\end{figure}

We use the normalized average hourly consumption over the day in December and January as clustering variables (i.e. 48 attributes) for each customer. $K$-means and hierarchical clustering are chosen and two cluster validity indexes SC and DB index are considered for the selection of optimal clustering algorithm and the number of clusters (a range between 2 and 8). Through the sub-clustering process for each tariff group, we obtained 13 sub-clusters in total. Due to the limitations of dataset (i.e. TOU tariffs remain the same), the merging of sub-clusters is implemented by grouping closest sub-cluster centroids through $K$-means with the desired final number of clusters as four. The results of adaptive customer segmentation are summarized in Table \ref{tab:cluster_summary}. In addition, the normalized electricity consumption profiles of four clusters are illustrated in the Figure \ref{fig:finalclusterconsumption} whereas corresponding cluster centroids are presented in Figure \ref{fig:clustercentroids}. It can be seen that each cluster has distinct characteristic such as different number of electricity consumption peaks and timing of peaks. In additional, electricity consumption behaviour changes over December 2009 and January 2010 could also be identified (e.g. see Figure \ref{fig:clustercentroids}) , which might be explained by the price effects (i.e. flat prices in December and TOU in January). 

\begin{table}[!t]
	\caption{Summary of results of the adaptive clustering based customer segmentation. }
	\centering
	\begin{tabular}{ll|l|l|l|l|ll}
		\hline
		\multicolumn{2}{c|}{Initial groups}                                                              & \multicolumn{4}{c|}{Sub-clustering stage}                                                                               & \multicolumn{2}{c}{Final merging stage}                                                         \\ \hline
		\multicolumn{1}{l|}{Group } & \begin{tabular}[c]{@{}l@{}}Size of \\ each group\end{tabular} & Group name & \begin{tabular}[c]{@{}l@{}}Optimal algorithms \\ and number of \\ sub-clusters\end{tabular} & SC   & DB index & \multicolumn{1}{l|}{Group } & \begin{tabular}[c]{@{}l@{}}Size of \\ each group\end{tabular} \\ \hline
		\multicolumn{1}{l|}{TA}         &          1156                                                     & TA         & K-means/ 3                                                                               & 0.18 & 2.27     & \multicolumn{1}{l|}{Cluster 1}  &    750                                                           \\ \hline
		\multicolumn{1}{l|}{TB}         &     438                                                          & TB         & K-means/ 3                                                                               & 0.19 & 2.29     & \multicolumn{1}{l|}{Cluster 2}  &     1154                                                          \\ \hline
		\multicolumn{1}{l|}{TC}         &      1181                                                         & TC         & K-means/ 3                                                                               & 0.18 & 2.31     & \multicolumn{1}{l|}{Cluster 3}  &     70                                                          \\ \hline
		\multicolumn{1}{l|}{TD}         &       433                                                        & TD         & K-means/ 4                                                                               & 0.15 & 2.37     & \multicolumn{1}{l|}{Cluster 4}  &           1234                                                    \\ \hline
	\end{tabular}
	\label{tab:cluster_summary}
\end{table}

\subsection{Multiple Pricing Optimization}

\highlightnew{We evaluate our multiple pricing optimization as follows: first, we identify customized demand models for each group of customers and the aggregated demand model for all customers. Based on the aggregated demand model for all customers, we use the uniform pricing optimization version of the proposed pricing model to obtain optimal uniform prices for all customers over the optimization horizon (i.e. all customers receive the same pricing decision); secondly, based on identified customized demand models for each customer groups/clusters, we adopt multiple pricing optimization to find optimal multiple pricing strategies for different types of customers (i.e. different types of customers receive different pricing decisions). }

\subsubsection{Group Demand Modelling}

Due to the lack of individual-level demand and price data, we build group demand models based on modified wholesale electricity market data. It should be mentioned that there are emerging interests from academia and industry in enabling open energy data such as individual-level data for research and development purpose. Therefore, it is credible to predict such household level data will become publicly available in the foreseeable future.  

Motivated by \cite{kirschen2000factoring}, we consider a residential district with 3000 households comprised of three types of customers: 1000 insensitive households who are typically insensitive to price changes (IS customers), 1000 curtailable customers who will reduce their electricity usage at some hours when corresponding prices are high but less willing to shift their electricity usage (SC customers) and 1000 shiftable and curtailable customers who are willing to shift their electricity usage from peak hours to off-peak hours and curtail their electricity usage at high price periods (SCS customers). 

First, the PJM day-ahead energy market demand and price data for the period between December 20, 2018 and December 19, 2020 \cite{pjm2019} are downscaled for the group demand modelling of IS customers. Second, demand models of SC and SCS customers are modified directly from identified demand models of IS customers to reflect SC customers have  higher absolute values of self-elasticity  and SCS customers have higher absolute values of both self and cross elasticities.

\begin{figure}[!ht]
	\centering
	\includegraphics[width=12cm]{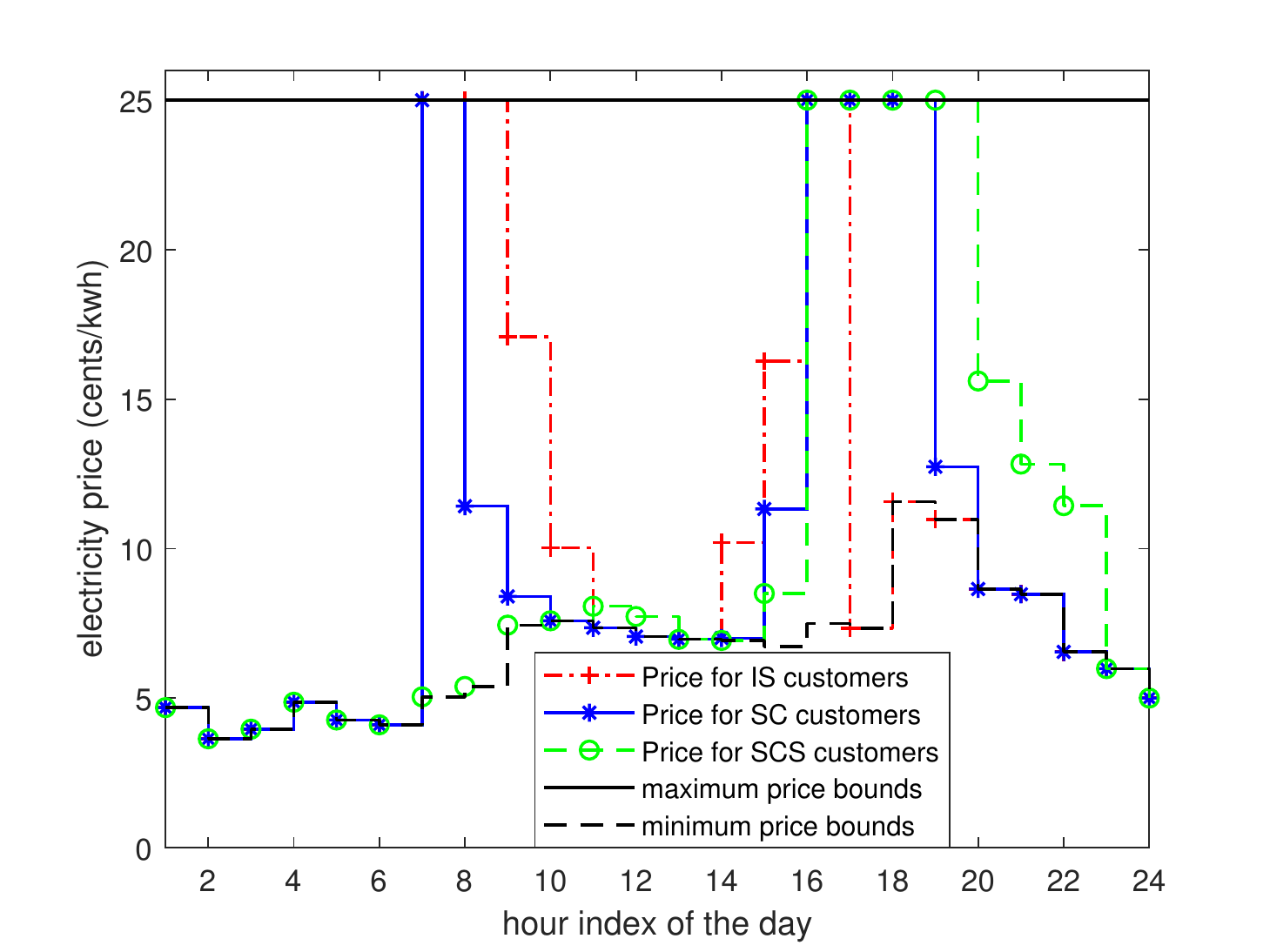}  
	\caption{Optimal multiple prices for different customer groups.}
	\label{fig:multiple_price}
\end{figure}
\begin{figure}[!ht]
	\centering
	\includegraphics[width=12cm]{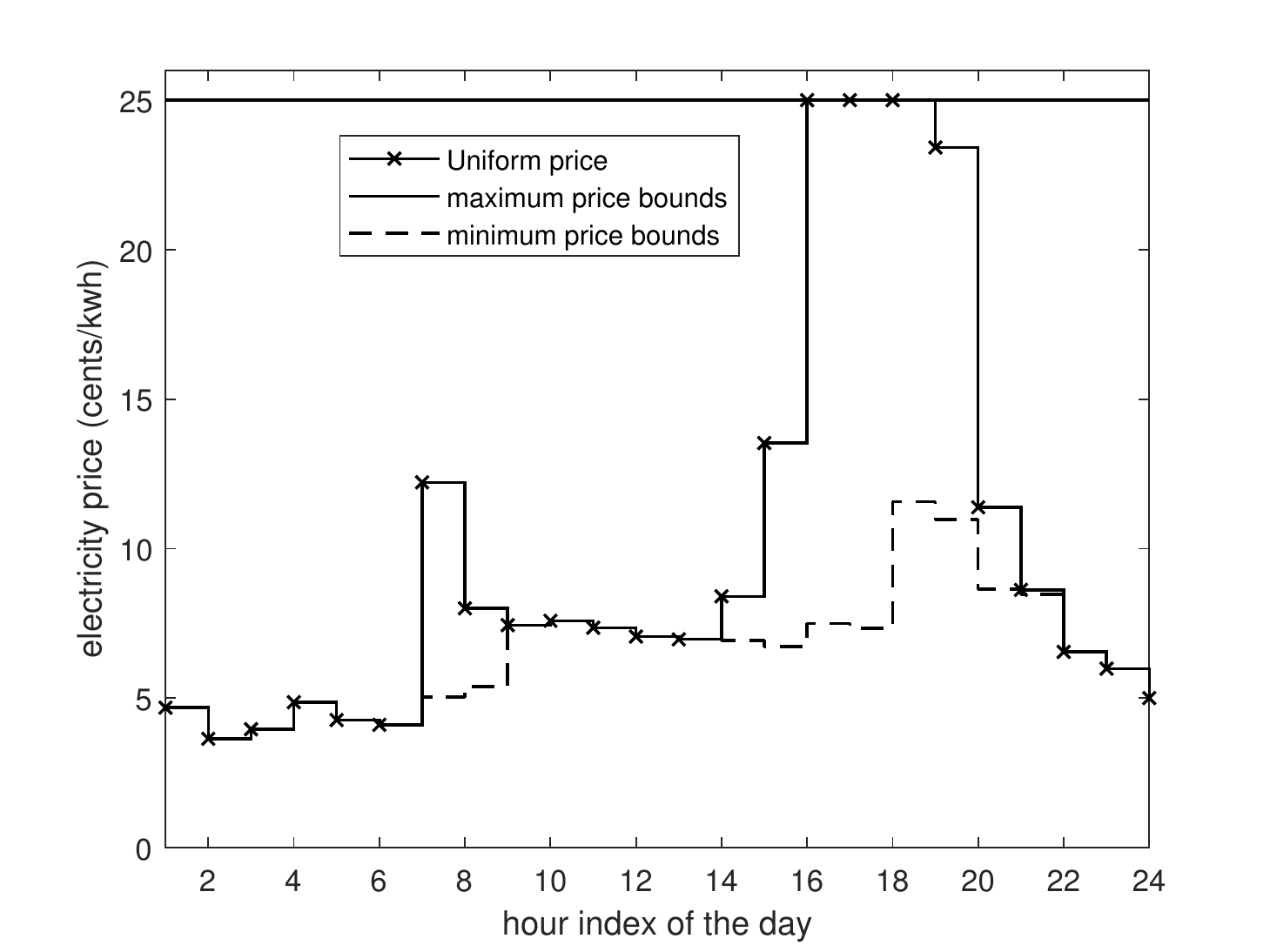}  
	\caption{Optimal uniform price for all customers.}
	\label{fig:uniform_price}
\end{figure}

\subsubsection{Multiple Pricing Optimization Results} For the pricing optimization, the parameters are set as follows: the maximum and minimum price bounds are set to 25 cents and the purchasing cost (wholesale price) respectively whereas the average price is set to 10 cents. The obtained optimal multiple prices and uniform prices are presented in Figures \ref{fig:multiple_price} and \ref{fig:uniform_price} respectively.  It can be seen that different from the uniform pricing, the multiple pricing has the flexibility to offer customized prices to different types of customers. In particular, prices for SCS customers are fluctuated across the day, which provides the opportunity for customers to make energy shifting decisions. In comparison, prices for SC customers are relatively stable with two peak price periods around the morning and evening. This is due to the fact that SC customers will usually curtail their usage during high price time periods but have less flexibility in energy shifting. For IS customers, they have a morning peak price and interestingly an early evening peak price (4-5pm). This could be explained by the existence of the fixed average price constraint in the pricing optimization model such that excessive bills can be avoided for such group of customers. To compare the profit gained by the retailer through two different pricing strategies, we re-run the optimization for another nine times using simulated electricity purchase cost data. The results show that compared with uniform pricing, the retailer could achieve around 2.1\% improvement in profit on average. 

The above simulation results give promising indications: 1) the proposed pricing method is able to provide the important flexibility such that right prices are offered to right customers; 2) the retailer has  intrinsic incentives to adopt the proposed multiple pricing approach; 3) By properly implemented, the proposed multiple pricing framework has potential to achieve benefits for different stakeholders in the system.

\section{Conclusion} \label{conclusion}
In this paper, we propose a multiple dynamic pricing framework for the smart grid retail market, which consists of key components of adaptive clustering-based customer segmentation, customized demand modelling, and multiple pricing optimization. First, an adaptive clustering based customer segmentation is implemented to obtain desirable customer segments. Second, customized demand models are learned for each group of customers in a data-driven manner with important and realistic market constraints included. Compared with the uniform pricing (low flexibility in offering customized services) and individual pricing (high randomness and poor scalability), by considering different usage patterns among customers, the multiple pricing approach enabled by the dynamic and adaptive customer segmentation and customized demand modelling can not only offer right prices to right customers but can also enable scalable and practical applications for the retailer with many customers. Finally, the effectiveness of the proposed framework is evaluated through simulations supported by the real-world data. 

\bibliographystyle{model5-names}
\biboptions{authoryear}
\bibliography{multiple_pricing.bib}

\begin{thebibliography}{27}
\expandafter\ifx\csname natexlab\endcsname\relax\def\natexlab#1{#1}\fi
\providecommand{\url}[1]{\texttt{#1}}
\providecommand{\href}[2]{#2}
\providecommand{\path}[1]{#1}
\providecommand{\DOIprefix}{doi:}
\providecommand{\ArXivprefix}{arXiv:}
\providecommand{\URLprefix}{URL: }
\providecommand{\Pubmedprefix}{pmid:}
\providecommand{\doi}[1]{\href{http://dx.doi.org/#1}{\path{#1}}}
\providecommand{\Pubmed}[1]{\href{pmid:#1}{\path{#1}}}
\providecommand{\bibinfo}[2]{#2}
\ifx\xfnm\relax \def\xfnm[#1]{\unskip,\space#1}\fi
\bibitem[{Archive(2009)}]{cer_smart_meter}
\bibinfo{author}{Archive, I. S. S.~D.} (\bibinfo{year}{2009}).
\newblock \bibinfo{title}{Cer smart metering project}.
\newblock \bibinfo{howpublished}{http://www.ucd.ie/issda/}.
\newblock \bibinfo{note}{Accessed 15 January 2020}.
\bibitem[{Asadinejad et~al.(2016)Asadinejad, Varzaneh, Tomsovic, Chen \&
  Sawhney}]{asadinejad2016residential}
\bibinfo{author}{Asadinejad, A.}, \bibinfo{author}{Varzaneh, M.~G.},
  \bibinfo{author}{Tomsovic, K.}, \bibinfo{author}{Chen, C.-f.}, \&
  \bibinfo{author}{Sawhney, R.} (\bibinfo{year}{2016}).
\newblock \bibinfo{title}{Residential customers elasticity estimation and
  clustering based on their contribution at incentive based demand response}.
\newblock In {\it \bibinfo{booktitle}{Power and Energy Society General Meeting
  (PESGM), 2016}\/} (pp. \bibinfo{pages}{1--5}).
\newblock \bibinfo{organization}{IEEE}.
\bibitem[{Bahrami et~al.(2017)Bahrami, Wong \& Huang}]{bahrami2017online}
\bibinfo{author}{Bahrami, S.}, \bibinfo{author}{Wong, V.~W.}, \&
  \bibinfo{author}{Huang, J.} (\bibinfo{year}{2017}).
\newblock \bibinfo{title}{An online learning algorithm for demand response in
  smart grid}.
\newblock {\it \bibinfo{journal}{IEEE Transactions on Smart Grid}\/},  {\it
  \bibinfo{volume}{9}\/}, \bibinfo{pages}{4712--4725}.
\bibitem[{Batista \& Batista(2018)}]{batista2018demand}
\bibinfo{author}{Batista, A.~C.}, \& \bibinfo{author}{Batista, L.~S.}
  (\bibinfo{year}{2018}).
\newblock \bibinfo{title}{Demand side management using a multi-criteria
  $\epsilon$-constraint based exact approach}.
\newblock {\it \bibinfo{journal}{Expert Systems with Applications}\/},  {\it
  \bibinfo{volume}{99}\/}, \bibinfo{pages}{180--192}.
\bibitem[{Chrysopoulos \& Mitkas(2017)}]{chrysopoulos2017customized}
\bibinfo{author}{Chrysopoulos, A.}, \& \bibinfo{author}{Mitkas, P.}
  (\bibinfo{year}{2017}).
\newblock \bibinfo{title}{Customized time-of-use pricing for small-scale
  consumers using multi-objective particle swarm optimization}.
\newblock {\it \bibinfo{journal}{Advances in Building Energy Research}\/},
  (pp. \bibinfo{pages}{1--23}).
\bibitem[{Flath et~al.(2012)Flath, Nicolay, Conte, Dinther \&
  Filipova-Neumann}]{flath2012cluster}
\bibinfo{author}{Flath, C.}, \bibinfo{author}{Nicolay, D.},
  \bibinfo{author}{Conte, T.}, \bibinfo{author}{Dinther, C.}, \&
  \bibinfo{author}{Filipova-Neumann, L.} (\bibinfo{year}{2012}).
\newblock \bibinfo{title}{Cluster analysis of smart metering data}.
\newblock {\it \bibinfo{journal}{Business \& Information Systems Engineering:
  The International Journal of WIRTSCHAFTSINFORMATIK}\/},  {\it
  \bibinfo{volume}{4}\/}, \bibinfo{pages}{31--39}.
\bibitem[{Haben et~al.(2016)Haben, Singleton \& Grindrod}]{haben2016analysis}
\bibinfo{author}{Haben, S.}, \bibinfo{author}{Singleton, C.}, \&
  \bibinfo{author}{Grindrod, P.} (\bibinfo{year}{2016}).
\newblock \bibinfo{title}{Analysis and clustering of residential customers
  energy behavioral demand using smart meter data}.
\newblock {\it \bibinfo{journal}{IEEE Transactions on Smart Grid}\/},  {\it
  \bibinfo{volume}{7}\/}, \bibinfo{pages}{136--144}.
\bibitem[{Hayes et~al.(2017)Hayes, Melatti, Mancini, Prodanovic \&
  Tronci}]{hayes2017residential}
\bibinfo{author}{Hayes, B.}, \bibinfo{author}{Melatti, I.},
  \bibinfo{author}{Mancini, T.}, \bibinfo{author}{Prodanovic, M.}, \&
  \bibinfo{author}{Tronci, E.} (\bibinfo{year}{2017}).
\newblock \bibinfo{title}{Residential demand management using individualized
  demand aware price policies}.
\newblock {\it \bibinfo{journal}{IEEE Transactions on Smart Grid}\/},  {\it
  \bibinfo{volume}{8}\/}, \bibinfo{pages}{1284--1294}.
\bibitem[{Kirschen et~al.(2000)Kirschen, Strbac, Cumperayot \&
  de~Paiva~Mendes}]{kirschen2000factoring}
\bibinfo{author}{Kirschen, D.~S.}, \bibinfo{author}{Strbac, G.},
  \bibinfo{author}{Cumperayot, P.}, \& \bibinfo{author}{de~Paiva~Mendes, D.}
  (\bibinfo{year}{2000}).
\newblock \bibinfo{title}{Factoring the elasticity of demand in electricity
  prices}.
\newblock {\it \bibinfo{journal}{IEEE Transactions on Power Systems}\/},  {\it
  \bibinfo{volume}{15}\/}, \bibinfo{pages}{612--617}.
\bibitem[{Kotouza et~al.(2017)Kotouza, Chrysopoulos \&
  Mitkas}]{kotouza2017segmentation}
\bibinfo{author}{Kotouza, M.~T.}, \bibinfo{author}{Chrysopoulos, A.~C.}, \&
  \bibinfo{author}{Mitkas, P.~A.} (\bibinfo{year}{2017}).
\newblock \bibinfo{title}{Segmentation of low voltage consumers for designing
  individualized pricing policies}.
\newblock In {\it \bibinfo{booktitle}{European Energy Market (EEM), 2017 14th
  International Conference on the}\/} (pp. \bibinfo{pages}{1--6}).
\newblock \bibinfo{organization}{IEEE}.
\bibitem[{Lu et~al.(2018)Lu, Hong \& Zhang}]{lu2018dynamic}
\bibinfo{author}{Lu, R.}, \bibinfo{author}{Hong, S.~H.}, \&
  \bibinfo{author}{Zhang, X.} (\bibinfo{year}{2018}).
\newblock \bibinfo{title}{A dynamic pricing demand response algorithm for smart
  grid: Reinforcement learning approach}.
\newblock {\it \bibinfo{journal}{Applied Energy}\/},  {\it
  \bibinfo{volume}{220}\/}, \bibinfo{pages}{220--230}.
\bibitem[{Martinez-Pabon et~al.(2018)Martinez-Pabon, Eveleigh \&
  Tanju}]{martinez2018optimizing}
\bibinfo{author}{Martinez-Pabon, M.}, \bibinfo{author}{Eveleigh, T.}, \&
  \bibinfo{author}{Tanju, B.} (\bibinfo{year}{2018}).
\newblock \bibinfo{title}{Optimizing residential energy management using an
  autonomous scheduler system}.
\newblock {\it \bibinfo{journal}{Expert Systems with Applications}\/},  {\it
  \bibinfo{volume}{96}\/}, \bibinfo{pages}{373--387}.
\bibitem[{Meng et~al.(2017)Meng, Kazemtabrizi, Zeng \& Dent}]{meng2017optimal}
\bibinfo{author}{Meng, F.}, \bibinfo{author}{Kazemtabrizi, B.},
  \bibinfo{author}{Zeng, X.-J.}, \& \bibinfo{author}{Dent, C.}
  (\bibinfo{year}{2017}).
\newblock \bibinfo{title}{An optimal differential pricing in smart grid based
  on customer segmentation}.
\newblock In {\it \bibinfo{booktitle}{Innovative Smart Grid Technologies
  Conference Europe (ISGT-Europe), 2017 IEEE PES}\/} (pp.
  \bibinfo{pages}{1--6}).
\newblock \bibinfo{organization}{IEEE}.
\bibitem[{PJM(2018)}]{pjm2019}
\bibinfo{author}{PJM} (\bibinfo{year}{2018}).
\newblock \bibinfo{title}{Day-ahead energy market}.
\newblock
  \bibinfo{howpublished}{https://www.pjm.com/markets-and-operations/energy}.
\newblock \bibinfo{note}{Accessed 15 January 2020}.
\bibitem[{Simshauser \& Whish-Wilson(2017)}]{simshauser2017price}
\bibinfo{author}{Simshauser, P.}, \& \bibinfo{author}{Whish-Wilson, P.}
  (\bibinfo{year}{2017}).
\newblock \bibinfo{title}{Price discrimination in australia's retail
  electricity markets: An analysis of victoria \& southeast queensland}.
\newblock {\it \bibinfo{journal}{Energy Economics}\/},  {\it
  \bibinfo{volume}{62}\/}, \bibinfo{pages}{92--103}.
\bibitem[{Srinivasan et~al.(2017)Srinivasan, Rajgarhia, Radhakrishnan, Sharma
  \& Khincha}]{srinivasan2017game}
\bibinfo{author}{Srinivasan, D.}, \bibinfo{author}{Rajgarhia, S.},
  \bibinfo{author}{Radhakrishnan, B.~M.}, \bibinfo{author}{Sharma, A.}, \&
  \bibinfo{author}{Khincha, H.} (\bibinfo{year}{2017}).
\newblock \bibinfo{title}{Game-theory based dynamic pricing strategies for
  demand side management in smart grids}.
\newblock {\it \bibinfo{journal}{Energy}\/},  {\it \bibinfo{volume}{126}\/},
  \bibinfo{pages}{132--143}.
\bibitem[{Tao et~al.(2017)Tao, Kun, Antti, Bjorn, Pertti \&
  Wencong}]{chen2017Classification}
\bibinfo{author}{Tao, C.}, \bibinfo{author}{Kun, Q.}, \bibinfo{author}{Antti,
  M.}, \bibinfo{author}{Bjorn, S.}, \bibinfo{author}{Pertti, J.}, \&
  \bibinfo{author}{Wencong, S.} (\bibinfo{year}{2017}).
\newblock \bibinfo{title}{Classification of electricity customer groups towards
  individualized price scheme design}.
\newblock In {\it \bibinfo{booktitle}{North American Power Symposium (NAPS),
  2017 IEEE}\/} (pp. \bibinfo{pages}{1--4}).
\newblock \bibinfo{organization}{IEEE}.
\bibitem[{Tushar et~al.(2017)Tushar, Yuen, Smith \& Poor}]{tushar2017price}
\bibinfo{author}{Tushar, W.}, \bibinfo{author}{Yuen, C.},
  \bibinfo{author}{Smith, D.~B.}, \& \bibinfo{author}{Poor, H.~V.}
  (\bibinfo{year}{2017}).
\newblock \bibinfo{title}{Price discrimination for energy trading in smart
  grid: A game theoretic approach}.
\newblock {\it \bibinfo{journal}{IEEE Transactions on Smart Grid}\/},  {\it
  \bibinfo{volume}{8}\/}, \bibinfo{pages}{1790--1801}.
\bibitem[{Wang et~al.(2020)Wang, Li, Ming \& Wang}]{wang2020deep}
\bibinfo{author}{Wang, B.}, \bibinfo{author}{Li, Y.}, \bibinfo{author}{Ming,
  W.}, \& \bibinfo{author}{Wang, S.} (\bibinfo{year}{2020}).
\newblock \bibinfo{title}{Deep reinforcement learning method for demand
  response management of interruptible load}.
\newblock {\it \bibinfo{journal}{IEEE Transactions on Smart Grid}\/},  {\it
  \bibinfo{volume}{11}\/}, \bibinfo{pages}{3146--3155}.
\bibitem[{Ward~Jr(1963)}]{ward1963hierarchical}
\bibinfo{author}{Ward~Jr, J.~H.} (\bibinfo{year}{1963}).
\newblock \bibinfo{title}{Hierarchical grouping to optimize an objective
  function}.
\newblock {\it \bibinfo{journal}{Journal of the American statistical
  association}\/},  {\it \bibinfo{volume}{58}\/}, \bibinfo{pages}{236--244}.
\bibitem[{Wilson(1993)}]{wilson1993nonlinear}
\bibinfo{author}{Wilson, R.~B.} (\bibinfo{year}{1993}).
\newblock {\it \bibinfo{title}{Nonlinear pricing}\/}.
\newblock \bibinfo{publisher}{Oxford University Press}.
\bibitem[{Yang et~al.(2018{\natexlab{a}})Yang, Zhao, Wen \&
  Dong}]{yang2018model}
\bibinfo{author}{Yang, J.}, \bibinfo{author}{Zhao, J.}, \bibinfo{author}{Wen,
  F.}, \& \bibinfo{author}{Dong, Z.} (\bibinfo{year}{2018}{\natexlab{a}}).
\newblock \bibinfo{title}{A model of customizing electricity retail prices
  based on load profile clustering analysis}.
\newblock {\it \bibinfo{journal}{IEEE Transactions on Smart Grid}\/},  {\it
  \bibinfo{volume}{10}\/}, \bibinfo{pages}{3374--3386}.
\bibitem[{Yang et~al.(2018{\natexlab{b}})Yang, Zhao, Wen \&
  Dong}]{yang2018framework}
\bibinfo{author}{Yang, J.}, \bibinfo{author}{Zhao, J.}, \bibinfo{author}{Wen,
  F.}, \& \bibinfo{author}{Dong, Z.~Y.} (\bibinfo{year}{2018}{\natexlab{b}}).
\newblock \bibinfo{title}{A framework of customizing electricity retail
  prices}.
\newblock {\it \bibinfo{journal}{IEEE Transactions on Power Systems}\/},  {\it
  \bibinfo{volume}{33}\/}, \bibinfo{pages}{2415--2428}.
\bibitem[{de~Zepeda et~al.(2021)de~Zepeda, Meng, Su, Zeng \&
  Wang}]{de97dynamic}
\bibinfo{author}{de~Zepeda, M. V.~N.}, \bibinfo{author}{Meng, F.},
  \bibinfo{author}{Su, J.}, \bibinfo{author}{Zeng, X.-J.}, \&
  \bibinfo{author}{Wang, Q.} (\bibinfo{year}{2021}).
\newblock \bibinfo{title}{Dynamic clustering analysis for driving styles
  identification}.
\newblock {\it \bibinfo{journal}{Engineering Applications of Artificial
  Intelligence}\/},  {\it \bibinfo{volume}{97}\/}, \bibinfo{pages}{104096}.
\bibitem[{Zhang et~al.(2017)Zhang, Bao, Yu, Yang \& Han}]{zhang2017deep}
\bibinfo{author}{Zhang, X.}, \bibinfo{author}{Bao, T.}, \bibinfo{author}{Yu,
  T.}, \bibinfo{author}{Yang, B.}, \& \bibinfo{author}{Han, C.}
  (\bibinfo{year}{2017}).
\newblock \bibinfo{title}{Deep transfer q-learning with virtual leader-follower
  for supply-demand stackelberg game of smart grid}.
\newblock {\it \bibinfo{journal}{Energy}\/},  {\it \bibinfo{volume}{133}\/},
  \bibinfo{pages}{348--365}.
\bibitem[{Zhong et~al.(2021)Zhong, Wang, Zhao, Li, Li, Wang, Deng \&
  Zhu}]{zhong2021deep}
\bibinfo{author}{Zhong, S.}, \bibinfo{author}{Wang, X.}, \bibinfo{author}{Zhao,
  J.}, \bibinfo{author}{Li, W.}, \bibinfo{author}{Li, H.},
  \bibinfo{author}{Wang, Y.}, \bibinfo{author}{Deng, S.}, \&
  \bibinfo{author}{Zhu, J.} (\bibinfo{year}{2021}).
\newblock \bibinfo{title}{Deep reinforcement learning framework for dynamic
  pricing demand response of regenerative electric heating}.
\newblock {\it \bibinfo{journal}{Applied Energy}\/},  {\it
  \bibinfo{volume}{288}\/}, \bibinfo{pages}{116623}.
\bibitem[{Zugno et~al.(2013)Zugno, Morales, Pinson \&
  Madsen}]{zugno2013bilevel}
\bibinfo{author}{Zugno, M.}, \bibinfo{author}{Morales, J.~M.},
  \bibinfo{author}{Pinson, P.}, \& \bibinfo{author}{Madsen, H.}
  (\bibinfo{year}{2013}).
\newblock \bibinfo{title}{A bilevel model for electricity retailers'
  participation in a demand response market environment}.
\newblock {\it \bibinfo{journal}{Energy Economics}\/},  {\it
  \bibinfo{volume}{36}\/}, \bibinfo{pages}{182--197}.

\end{thebibliography}

\end{document}